\documentclass{PoS}

\usepackage{mathptmx}
\usepackage{fontenc}
\usepackage{times}
\usepackage{amssymb}
\usepackage{graphicx}

\usepackage{amsmath}
\usepackage{amsfonts}
\usepackage{bbold}
\usepackage{cite}
\usepackage{inputenc}

\newcommand{\al}{\alpha}
\newcommand{\be}{\beta}

\def\5{\overline 5}

\newcommand{\diag}{\text{diag}}

\newcommand{\unity}{\mathrm{Id}}

\def\eq#1{{(\ref{#1})}}

\def\Im{\mbox{Im}\,}

\def\hbar{\hspace{0pt}\raisebox{1pt}{$-$} \hspace{-7pt} h}

\newcommand{\beq}{\begin{equation}}
\newcommand{\eeq}{\end{equation}}
\newcommand{\bac}{\beq\begin{array}}
\newcommand{\eac}{\end{array}\eeq}
\newcommand{\ba}{\begin{array}}
\newcommand{\ea}{\end{array}}
\newcommand{\bea}{\begin{eqnarray}}
\newcommand{\eea}{\end{eqnarray}}
\title{Leptogenesis constraints from flavour symmetry induced Lepton Mixing}

\ShortTitle{Leptogenesis constraints from flavour symmetry induced Lepton
Mixing}

\author{\speaker{Ivo de Medeiros Varzielas}\\
        CFTP, Instituto Superior T\'ecnico\\
        E-mail: \email{ivo@cftp.ist.utl.pt}}

\abstract{In models with flavour symmetries added to the gauge group of the
Standard Model the CP-violating asymmetry necessary for leptogenesis may be
related with low-energy parameters. A particular case of interest is when the
flavour symmetry produces an exact mass independent lepton mixing scheme,
leading to a vanishing CP-violating asymmetry. We present a model-independent
discussion that confirms this always occurs for unflavoured leptogenesis in type
I see-saw scenarios.}

\FullConference{35th International Conference of High Energy Physics\\
		 July 22-28, 2010\\
		 Paris, France}

\begin{document}

\section{Introduction}

This submission is based
on \cite{ABdMM} where a complete discussion with references is presented.



The smallness of neutrino masses can be
understood within the see-saw mechanism where
the Standard Model (SM) is extended by adding new heavy states. In type I
see-saw the extra states are right-handed (RH)
neutrinos with large Majorana masses, and the light neutrino masses are
generated via effective operators with the effective matrix
given by
$m_\nu=-m_D \, M_R^{-1}\, m_D^T \,$.
In general $m_D$ and $M_R$ ($M_R=M_R^T$) are complex matrices.
We have
$\hat{m}_D=U_L^\dag  \, m_D\, U_R$ and $\hat{M}_R=V_R^T \, M_R\,V_R$, where the
``$\hat{\phantom{b}}$'' denotes a diagonal matrix and $U_L,U_R,V_R$ are the
diagonalising matrices.
The unitary matrices $U_\ell$, $U_{\ell^c}$ and $U_\nu$ diagonalise the charged
lepton and neutrino mass matrices and the lepton mixing
matrix is defined as $U=(U_\ell)^{\dagger} U_\nu$.
The leptonic mixing indicated by data is approximated by the
Tri-Bimaximal (TB) scheme, and this can be interpreted as a signal of an
underlying flavour symmetry. If $m_\nu$ is
exactly diagonalised by the TB mixing matrix $U_{TB}$ (in the basis with
diagonal charged lepton mass matrix) we write
$\hat{m}_\nu=D\,U_{TB}^T\, m_\nu\,U_{TB}\,D$, where
$D = \diag (e^{i \varphi_1}, e^{i\varphi_2}, 1)\,$ accounts for the low-energy
Majorana phases.
The requirement of having exact TB diagonalisation can be expressed
by rewriting the effective matrix as
\beq
\label{ssdiag}
\hat{m}_\nu= -  D\,(U_{TB}^T U_L)\,\hat{m}_D \, (U_R^\dag\,V_R)\,
\hat{M}_R^{-1}\, (V_R^T U_R^*)\,\hat{m}_D\,(U_L^T U_{TB})\,D\,.
\eeq


The see-saw mechanism also contains the ingredients for leptogenesis. We discuss
only ``unflavoured'' leptogenesis, as in
the framework of flavour symmetry models predicting TB mixing the
heavy singlet neutrinos typically have masses above $10^{13}$~GeV and
for $T\gtrsim 10^{12}$ GeV lepton flavours are indistinguishable.
In the standard thermal leptogenesis scenario singlet neutrinos $N_{\al}$ are
produced by scattering processes after inflation. Subsequent out-of-equilibrium
decays of these heavy states generate a CP-violating asymmetry given
by
\begin{equation}
  \label{eq:cp-asymm}
  \epsilon_{N_\al} = \frac{1}{4v^2 \pi (m_D^{R\,\dagger} \;m_D^R)_{\al\al}}
  \sum_{\be\neq \al} \Im \left[\left((m_D^{R\,\dagger}
\;m_D^R)_{\be\al}\right)^2\right]
  f(z_\be)\,,
\end{equation}
where $z_\be=M_\be^2/M_\al^2$, $f(z_\be)$ is the loop function,
$m_D^R \equiv m_D V_R$ (i.e. in the basis where $M_R$ is diagonal).



The orthogonal
complex matrix $R$ is defined by the Casas-Ibarra
parametrisation
\beq
  \label{eq:casas-ibarra}
R^*= (\hat{m}_\nu)^{-1/2} \,U^T\,m_D^R\, (\hat{M}_R)^{-1/2}\,.
\eeq
The matrix $R$ is very useful in expressing the CP-violating asymmetry.
Considering for simplicity hierarchical RH neutrinos ($M_1 \ll M_2 \ll M_3$),
with $m_j\equiv(\hat
m_\nu)_{jj}$, \eq{eq:cp-asymm} can
be rewritten as
\begin{equation}
  \label{eq:cp-asymm-CI}
  \epsilon_{N_\al} = -\frac{3 M_\al}{8 \pi v^2}
  \frac{ \Im
    \left[\sum_j m_j^2 R_{j\al}^2\right]}
  {\sum_j m_j |R_{j\al}|^2}\,.
\end{equation}
Fixing the RH neutrino mass spectrum and low-energy observables, random
values of $m^R_D$ correspond to random values of $R$.
Leptogenesis is insensitive to low-energy lepton mixing and its viability is not
related with any accidental mixing
pattern considered. The asymmetry in (\ref{eq:cp-asymm-CI}) is determined by
the values of the entries of $R$ which are arbitrary in the absence of any
flavour symmetry, therefore $\epsilon_{N_a}\neq 0$ in general.
\begin{figure}[t]
  \centering
  \includegraphics[width=4.5cm,height=3.25cm]{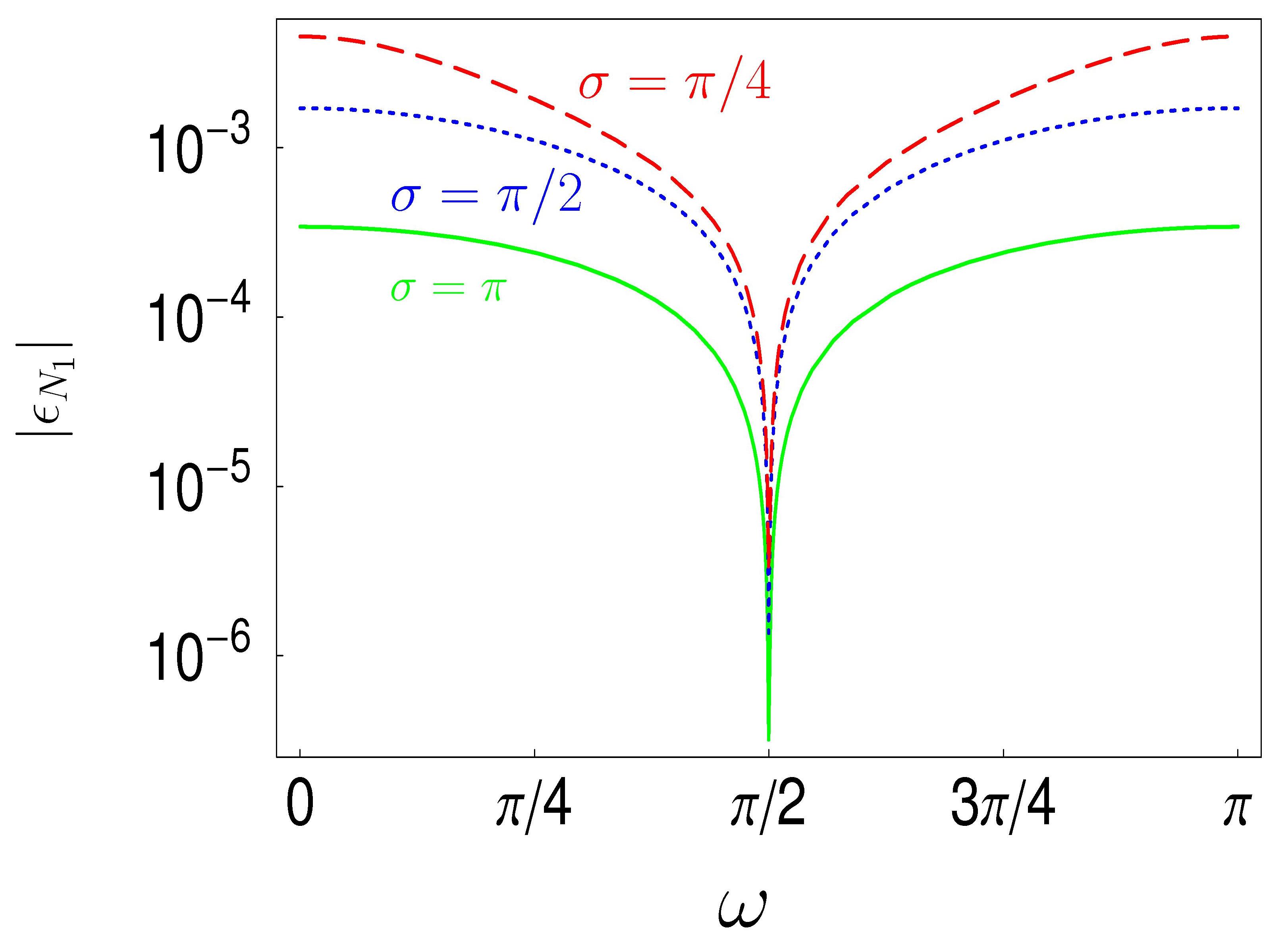}
  \caption{CP-violating asymmetry as a function of the angle $\omega$
    for different values of $\sigma$.}
  \label{fig:cp-asymm}
\end{figure}
To illustrate this, consider a
case where only $N_1$ decays are relevant and $R=R_{13}(\rho_{13})$
with $\rho_{13}=\omega + i
\sigma$ ($R$ is a $13$
rotation matrix with angle $\rho_{13}$) - Figure \ref{fig:cp-asymm} shows that
the asymmetry does not
vanish in general (regardless of any low-energy mixing pattern).

\section{Leptogenesis with form-diagonalisability}
\label{exTBFS}


We consider now that an underlying flavour symmetry
enforces an exact mixing pattern. In these cases, the mass matrices $m_D$ and
$M_R$ (controlled by the transformation properties of $L_i$ and $N_\alpha$
under the symmetry group) are form-diagonalisable, i.e. their eigenvalues are
completely
independent from their diagonalising
matrices, and the mixing matrix consists purely of
numbers. The following proof holds under form-diagonalisability, but we
assume TB mixing in particular (for definiteness).
Vanishing off-diagonal
elements of $\hat{m}_\nu$ in (\ref{ssdiag}) can arise only if
\begin{equation}
\label{eq:rot-mat-relations}
U_{TB}^T U_L= P_L \, O_{D_i} \quad \mbox{and}\quad
U_R^\dag\,V_R=O^\dag_{D_i}\,P_R\,O_{R_{m}}\,,
\end{equation}
where $P_{L,R}=\diag(e^{i\alpha^{R,L}_1},e^{i\alpha^{R,L}_2},
e^{i \alpha^{R,L}_3})$ whereas $O_{D_i}$ and $O_{R_m}$ are respectively unitary
and orthogonal matrices that arbitrarily rotate the
$i$ and $m$ degenerate eigenvalues of $m_D$ and $M_R$ such that if
$m_D$ (or $M_R$) has no degenerate eigenvalues $O_{D_i}= \diag(1,1,1) = \unity
\,$
(or $O_{R_m}=\unity$). The requirement of having canonical kinetic terms
in addition to preserving the $m$-fold degeneracy of the RH neutrino mass matrix
enforce $O_{R_m}$ to be real. Although $O_{D_i}$ and $O_{R_m}$ do not have any
effect in (\ref{ssdiag}) they affect the structure of
$U_{L,R}$ and $V_R$ and correspondingly of $m_D$. $V_R$ can be defined in such
a way that $\hat{M}_R$ is real, and
the phases contained in $\hat{m}_D$,
denoted by $\gamma_i$, must obey $\varphi_i + \alpha^{R}_i + \alpha^{L}_i +
\gamma_i= 2 k \pi$
and $\alpha^{R}_3 + \alpha^{L}_3 + \gamma_3 = 2 n \pi$. To understand the
conditions given in (\ref{eq:rot-mat-relations}) consider for
simplicity the case without any degeneracy: $O_{D_i}=\unity$ and
$O_{R_m}=\unity$. If the products $U_{TB}^TU_L$ and $U_R^\dag V_R$ are not
diagonal then the right-hand side of (\ref{ssdiag}) is in general a matrix with
entries that are
linear combinations of the mass eigenvalues of $\hat{m}_D$ and of $\hat{M}_R$.
In order to have the the off-diagonal entries of $\hat{m}_\nu$ vanish the
respective linear combinations must all cancel out, but such
cancellations correspond to having well-defined relationships between the
eigenvalues of
$\hat{m}_D$ and of $\hat{M}_R$ - it would require extreme fine-tuning of the
masses. In order to avoid that one can instead consider that $U_{TB}^TU_L$ and
$U_R^\dag V_R$ are diagonal (or they must obey (\ref{eq:rot-mat-relations}), in
the general case with degeneracies).



Taking $O_{D_i}=\unity$ and
$O_{R_m}=\unity$ for simplicity (the case with degeneracies is treated
carefully in \cite{ABdMM}), we
write $\hat{m}_D=\diag(v_1,v_2,v_3)$ in the $\hat{m}_D$ basis, and in
the $\hat{M}_R$ basis we have
\beq
\label{vCI}
m_D^R= U_{TB}\,P_L\, \diag (v_1,v_2,v_3)\,\,P_R\,,
\eeq
then redefining the $v_i$ by
absorbing $P_L,P_R$ such that $m^R_D= U_{TB}\, \diag (v_1,v_2,v_3)$ we
finally write
\beq
\label{vCI1}
m_D^R= U_{TB}\,P\, \hat{v}\, ,
\eeq
$\hat{v}= \diag(|v_1|,|v_2|,|v_3|)$ and the phases put into $P$.
The type I see-saw formula of (\ref{ssdiag}) becomes
\beq
\label{vCI2}
\hat{m}_\nu =- D\, U_{TB}^T\, ( U_{TB} \,P\, \hat{v}\,)\hat{M}_R^{-1}
(\hat{v} \,P\, U_{TB}^T) U_{TB} \,D =  (\hat{v} \hat{M}_R^{-1/2}
R^\dag)(R^*  \hat{M}_R^{-1/2} \hat{v} ).
\eeq
We used $D=P^*\, \, e^{-i \pi/2}$, which is a consequence of our definition of
$\hat{m}_\nu$ in  \eq{ssdiag}, and we introduced the arbitrary
orthogonal complex matrix  $R$.
From \eq{vCI2} we have that
\beq
\hat{m}_\nu^{-1/2} \, \hat{v} \,\hat{M}_R^{-1/2}  R^\dag= \unity \,.
\eeq
As $R^\dag R^*= R^T R=\unity$ we arrive at the following expression with $R$
related to diagonal matrices
\beq
\label{ourR}
R^*=\hat{m}_\nu^{-1/2}\,  \hat{v}\, \hat{M}_R^{-1/2} \,.
\eeq
By comparing \eq{ourR} with the parametrisation given in
\eq{eq:casas-ibarra} we deduce that in the case of exact TB mixing the
matrix $R$ is real and according to \eq{eq:cp-asymm-CI} the
CP-violating asymmetry vanishes.

It is straightforward to check that the matrix $R$ still turns out to be
real for other exact
mixing schemes as long as their mixing matrix also consists purely of numbers
(e.g. the Bi-maximal mixing scheme).
Note also that the result is also generalisable to models with two or
more than three  RH neutrinos.

\section{Conclusions}

An important consequence of our proof is that if the TB mixing
pattern is due to an underlying flavour symmetry in a type I see-saw
scenario, the viability of leptogenesis depends upon possible
departures from the exact pattern. In the context of models based on
discrete flavour symmetries that predict TB mixing at leading order this is
achieved through next to leading order corrections. Since the size of the
deviations from TB mixing are not arbitrary (constrained by data), one might
expect the CP-violating asymmetry to be constrained by low-energy
observables.

We refer the interested reader to \cite{ABdMM} for a more detailed discussion
of this proof (as well as other related topics not discussed here). Importantly,
\cite{ABdMM} includes also several relevant references that have been omitted
here due to the limited space.

\section*{Acknowledgements}

We thank Luca Merlo for useful discussions.
The work of IdMV was supported by FCT under the grant
SFRH/BPD/35919/2007.  The work of IdMV was partially supported by FCT
through the projects POCI/81919/2007, CERN/FP/83503/2008 and CFTP-FCT
UNIT 777 which are partially funded through POCTI (FEDER) and by the
Marie Curie RTN MRTN-CT-2006-035505.


\end{document}